\documentclass{llncs}
\usepackage{graphicx}
\usepackage{todonotes}
\usepackage{comment}
\usepackage{enumitem}
\usepackage{microtype}
\usepackage{setspace}
\usepackage[hyphens]{url}
\usepackage{hyperref}
\title{A Report on the Cost of Data Privacy}

\author{Monika Balamurugan}
\institute{Technische Universität Dortmund \\ 
\email{monika.balamurugan@tu-dortmund.de}}

\begin{document}
\maketitle
\begin{abstract}
As our lives migrate to the digital realm, our online identity has evolved to become an increasingly robust collection of data about every aspect of our online and offline lives. This data is extremely appealing to companies who wish to use it for a variety of analytics. In this report, we create awareness for the consumers around the world who have only a vague understanding of how much of their data is being tracked, where, when, and by which companies.


\end{abstract}

\section{Introduction} \label{Intro}

In today's digital era, companies are collecting data about our online activities, browsing the Internet, posting on social media, etc. Together, such collected data make up a digital identity for each person. There is no official list of what data is collected or leaves the consumer devices, they are also confused about where they are being tracked. Across the world, people are uncomfortable with how their data is being used by companies, even when it comes to worthy causes like COVID-19 tracking. We report that consumers often have to sacrifice their privacy to access their desired services, thus leading to decreasing levels of trust. There are \emph{3 Ts} namely Tracking, Trust, and Time that reveal the trade-offs faced by consumers when using online services. This report will summarise the findings of that survey. Our report is a summary of data from Cshub \footnote{https://www.cshub.com/whitepapers}.

\section{Awareness on Social Media Tracking}

As consumers scroll through their feeds, they need to have a general awareness that companies are tracking their data. Many believe that social media companies track at least one element of their personal data. People from the Netherlands likely expect social media companies to actively collect data on their posts and only 40\% do not think their data is safe. In France, 59\% of French people are not aware that their posts are being tracked by social media companies, and 45\% of Australians agree that their data is being tracked. 

There is a popular expression that \emph{if you’re not paying for the product, you are the product}. The Internet provides a wealth of information and free services to consumers. But people often forget that the free service they receive from an Internet search, and social media account are being funded, and advertisers are buying access to them. From a recent report from Cambridge Analytica, the My Health Record in Australia has suffered potential data breaches, shrinking levels of privacy. Yet, many consumers are not aware of the routine private data tracking and data harvesting that is common today. 

\section{Tracking a Global Pandemic like COVID-19}

Recently people have been immersed with COVID-19 news, which provides rapid updates about case counts, reopening plans, etc. Hence, just 4\% of Australians are unaware of the smartphone data collection efforts that aim to track the spread of COVID-19. In other countries like Dutch, 14\% of respondents are unaware, 35\% of Americans are unaware,  and 17\% of Germans are unaware of data tracking efforts to contain the novel coronavirus. In Australia, 19\% of people are opposed to data tracking even for the COVID-19 purpose. Hence COVID-19 is closing the data tracking education gap. Whether or not people understand which data is being tracked, they are overall unhappy about companies tracking their data, regardless of its purpose of collection. Around 91\% of people are uncomfortable with at least one kind of company tracking their data.

\section{Can Edge Analytics Protect Real-life Data}

In addition to the data collection concerns, people worry more if the data collected are used to directly make money. Around 93\% of people considered in the survey are uncomfortable with companies selling their data. Nine out of 10 Australians are upset with companies profiting using any of their collected personal data. More importantly, the sale of biometric data, passwords, and offline conversations cause the most concern at 86\% each. Hence, there is a strong need for analytics and data processing to be performed offline on the device owned by the consumers. i.e., data should not be transmitted out of the device even for authentication or for advanced analytics purposes. In the following, we brief some of the recent works that accomplish the tasks demanded by users, offline, at the device level.

Today's DL models when efficiently deployed can convert normal IoT devices into intelligent IoT devices that can solve a wide variety of problems. For example, in \cite{aics19smartspeaker}, a face recognition algorithm was trained using Deep Neural Network and deployed on their modern Alexa smart speaker prototype. This model, without disturbing the smart speaker routine, can detect and identify a human face and start the Alexa voice service only when an authorized face is present in the live video frames. Similarly, in \cite{aivision}, a DL and Open CV based object detection model was deployed in their smart speaker. Here, whenever the user calls out the command \emph{Alexa, ask Friday what she sees}, the smart speaker camera turns on and executes the deployed model and calls out the names of detected objects as a response to the user’s command. The models optimized using methods such as \emph{Edge2train}  \cite{edge2trainiot2020} and \emph{RCE-NN} \cite{rcenniot2020} can run on the IoT devices. Processing data at the edge level without depending on the cloud improves latency, reduces subscription \& cloud storage costs, processing requirements, and bandwidth requirements. It also can address privacy and security issues by avoiding the transmission of sensitive or identifiable data over the network. 

Hybrid or composite approaches involving conventional CV and DL should be used to take great advantage of the limited computing resource available at the edge. Such heterogeneous systems consist of a combination of multiple processors and chipsets. For example, in the Smart Hearing Aid prototype from \cite{smarthearingaid}, the users have integrated a DSP-based microphone array with a Linux SBC to perform edge level audio processing such as noise suppression, the direction of arrival estimation, etc., without depending on the internet. The IoT devices can be power efficient when the user assigns different workloads to the most efficient compute engine \cite{internetcomputing}. For Deep Learning (DL) use cases, the SIFT \cite{sift}, SURF \cite{surf}, BRIEF \cite{brief} feature descriptors need to be generally combined with traditional machine learning classification algorithms \cite{sptiot} \cite{classpercom21} such as SVMs \cite{covidaway2020}, SVRs \cite{wfiot2020}, Random Forests \cite{sramoptimized}, Decision Trees \cite{anomalies} to solve the CV problems. DL is sometimes overkilling when the given problem is simple and if it can be solved by CV techniques. Algorithms like pixel counting, SIFT, simple color thresholding are not class-specific. i.e., they are very general and perform the same for any number of images.

\section{Distrust of Social Media Providers}

Almost half of the people have doubts that their conversations and biometric data are being tracked by companies, although there is consensus as to which type of services/companies are collecting data. People from the Netherlands show the highest rate of discomfort since 90\% of people are saying they are upset with personal conversation tracking, and 88\% are uncomfortable with biometric data gathering. The majority of people from various age groups, for example, 43\% of Australians are totally not comfortable with any form of personal data collected for any purpose. The most common data that people think companies are tracking is location data. 57\% are uncomfortable with services tracking their location. Recently in \cite{nakashima_2018_ap}, Google was storing location data on smartphones even if the consumer had turned off location access. 

The companies that provide us free service own our data that is generated when using their services. To give consumers control over their data being provided to companies, recently, some studies \cite{posner_2018_on} have proposed the idea that the services that collect personal data should pay the data owner (source of data generation). Although there is a provision for people to get compensated for their data, still 37\% say no, and 27\% are unsure if a one-time payment is worth the data. But for specific types of data (critical ones), 76\% of people are unwilling to sell at least one type of their data at any price. Hence, consumers value their data more than earning money.

\section{Summary}

We believe it is important to balance privacy and innovation. Consumers deserve to have control over their data, but at the same time, data is often essential to building new technologies and serving the common good like accurate tracking the spread of COVID-19. To strike a balance, organizations around the globe must embrace transparency and help their customers understand how their data is being used.

\begin{spacing}{0.92}
\bibliography{aics-sample.bib}
\bibliographystyle{splncs03}
\end{spacing}

\end{document}